\UseRawInputEncoding

\documentclass[pdflatex,sn-mathphys]{sn-jnl}

\usepackage{amssymb}
\usepackage{hyperref}
\usepackage{graphicx}
\usepackage{dcolumn}
\usepackage{bm}
\usepackage[utf8]{inputenc}
\usepackage[T1]{fontenc}
\usepackage{float}

\usepackage{color,soul}
\usepackage{mathtools}
\usepackage{epsfig}
\usepackage{color}
\usepackage{url}
\usepackage{siunitx}
\usepackage{enumitem}
\usepackage{caption}
\usepackage{subcaption}
\jyear{2022}%

\theoremstyle{thmstyleone}%
\theoremstyle{thmstyletwo}%

\theoremstyle{thmstylethree}%

\begin{document}

\title[Simulation Package for Analysis of Multilayer Spintronic Devices]{A Comprehensive Simulation Package for Analysis of Multilayer Spintronic Devices}


\author*[1]{\fnm{Jakub} \sur{Mojsiejuk}}\email{mojsieju@agh.edu.pl}

\author[1]{\fnm{Sławomir} \sur{Ziętek}}
\author[1,2]{\fnm{Krzysztof} \sur{Grochot}}
\author[1]{\fnm{Witold} \sur{Skowroński}}
\author[1,2]{\fnm{Tomasz} \sur{Stobiecki}}

\affil[2]{\orgdiv{Institute of Electronics}, \orgname{AGH University of Science and Technology}, \orgaddress{\street{Al. Mickiewicza 30}, \city{Kraków}, \postcode{30-059}, \country{Poland}}}

\affil[2]{\orgdiv{Faculty of Physics and Applied Computer Science}, \orgname{AGH University of Science and Technology}, \orgaddress{\street{Al. Mickiewicza 30}, \city{Kraków}, \postcode{30-059}, \country{Poland}}}


\abstract{We present \textsc{cmtj} -- a comprehensive simulation package that allows large-scale macrospin simulations for a variety of multilayer spintronics devices. Apart from conventional static simulations, such as magnetoresistance and magnetisation hysteresis loops, \textsc{cmtj} implements a mathematical model of dynamic experimental techniques commonly used for spintronics devices characterisation, for instance: spin diode ferromagnetic resonance, pulse-induced microwave magnetometry, or harmonic Hall voltage measurements. We demonstrate the accuracy of the macrospin simulations on a variety of examples, accompanied by some experimental results.}

\keywords{macrospin, Landau–Lifshitz-Gilbert-Slonczewski, spintronics, numerical}



\maketitle

\section{Introduction}
Modern development of spintronic devices \cite{dieny_opportunities_2020} requires careful design and a time-consuming fabrication process, for which preparation is often carried out with the help of simulations. The use of magnetic materials and multilayer structures with certain material parameters enables the design of a spintronic device with optimal functionality, depending on the application: memory (high endurance, low energy writing) \cite{ikegawa_magnetoresistive_2020}, sensor (high sensitivity, low noise) \cite{xu_macro-spin_2017} or high frequency component (high quality factor, low energy consumption, and low noise) \cite{hirohata_review_2020}. 

As we climb from the macromagnetic models, steadily increasing the resolution of the phenomena, and finally reaching the \textit{ab initio} simulations, we do so with increasing computational cost. This computational cost is closely correlated with the number and complexity of the simulation parameters, which hampers the speed of prototyping and the design of the experiment. As a result, we are choosing between a slower, but accurate approach and a faster, but not as precise method. In this paper, we do not undertake to solve this dilemma, but introduce tools that bring flexibility into the process of making that decision.  

\begin{figure}[ht]
    \centering
    \includegraphics{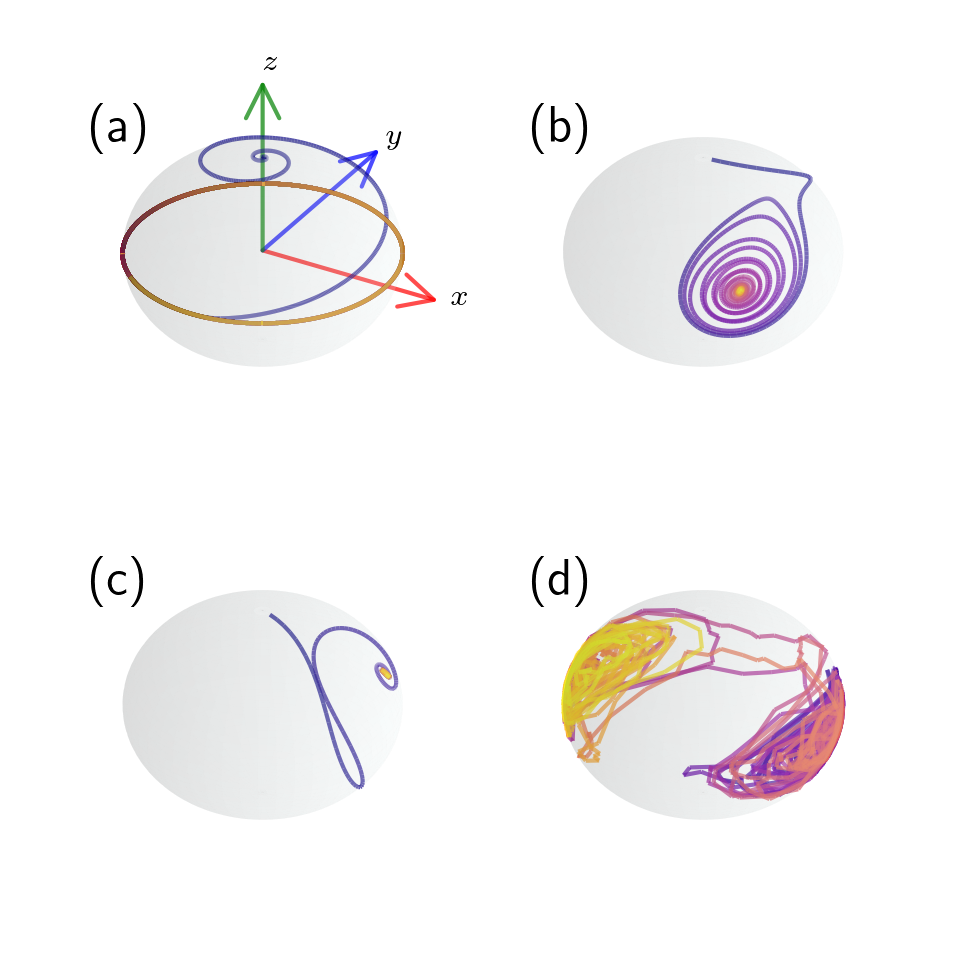}
    \caption{Example trajectories extracted by simulating a single ferromagnetic layer with different parameters and excitations. (a) depicts a stable STO (STT-excited system) trajectory under a constant current density. (b) demonstrates a trajectory obtained with exciting the magnetisation with VCMA. In (c) we see a trajectory under a pulse excitation of the Oersted field. Finally, (d) is a thin, bilayer ferromagnetic system with a low energy barrier that changes states from parallel to anti-parallel under thermal noise. All trajectories may be reproduced from the example codes available in documentation to \textsc{cmtj}.}
    \label{fig:traj}
\end{figure}

Micromagnetic packages, such as \textsc{MuMax3} \cite{vansteenkiste_design_2014} and \textsc{OOMMF} \cite{m._j._donahue_oommf_1999}, offer a lot of plasticity in modelling magnetic interactions of complex structures while maintaining an acceptable computational cost. On the other side of the spectrum, \textsc{VASP} \cite{kresse_ab_1993} and \textsc{Quantum Espresso} \cite{giannozzi_quantum_2009, giannozzi_advanced_2017} lead the way as acknowledged self-consistent solvers. In the macromagentic regime, a notable research direction was devoted to creating magnetic tunnel junction (MTJ) behavioural models in Verilog, for example, in the work of R. Garg et al. \cite{garg_behavioural_2012} or T. Chen et al. \cite{chen_comprehensive_2015, tingsu_chen_comprehensive_2015}. Such models attempt to capture the electronic nature of spintronics devices, describing them as discrete elements or electronic elements in a circuit, making it easy to prototype devices composed of discrete components. However, to our best knowledge, so far there have not been any wide-spread, easy-to-use, computationally efficient, open-sourced library that allows for quick testing and verification of ideas using macromagnetic simulations for magnetic multilayer systems including, but not limited to, static and dynamic characteristics of spin valves and magnetic tunnel junctions (MTJs). 
Thus, our main objective is to fill a gap in the magnetic simulation hierarchy by providing a standardised Python package, \textsc{cmtj}, for rapid prototyping and simulation. \textsc{cmtj} enables simulation of current-induced magnetisation dynamics calculations originating from spin transfer torque (STT) \cite{slonczewski_current-driven_1996,ogrodnik_study_2021} such as spin torque oscillator (STO) \cite{kiselev_microwave_2003}, STT-induced magnetisation switching \cite{myers_current-induced_1999} or voltage-controlled magnetic anisotropy (VCMA) phenomena \cite{nozaki_electric-field-induced_2012}. Some of those motivating examples are shown in Fig.\ref{fig:traj}.
In this work, we demonstrate a couple of standard use cases and workflows where \textsc{cmtj} provides valuable information, showing its ability to verify experimental setups. As an additional asset, we discuss \textsc{PyMag}, a graphical user interface (GUI) that is beneficial also for educational purposes, profiting from a user-friendly navigable front-end, but using a fast C++ back-end provided by \textsc{cmtj}.

First, we discuss the main theoretical components of the macromagnetic simulation. Then, we describe the structure of the simulation in the \textsc{cmtj} and critical software functionalities, such as software and driver systems. Finally, we discuss a couple of more advanced simulations that may be performed with \textsc{cmtj}. In the Appendix, we develop a short walk-through example for \textsc{PyMag}.

\section{\textsc{cmtj} backend}
The \textsc{cmtj} core is written in \textsc{C++} and provides a simple header-only library interface. This approach weighs on accessibility and compatibility with most platforms, as the user does not have to go through a complicated installation process. 
If the user wishes to benefit from the provided \textsc{python} interface, the setup involves a standard \textsc{pip} installation process. The scope of the \textsc{python} interface covers all basic functionalities of the \textsc{C++} core library, exposing the functions using the open source solution \textsc{Pybind11} \cite{noauthor_pybindpybind11_2022}. In addition, the \textsc{python} package offers utility functions that complement frequently used operations such as unit conversions, parallelism, parameter sweep, filtering, energy and resistance calculations, or template procedures for spin diode ferromagnetic resonance (SD-FMR) or pulse-induced microwave magnetometry (PIMM).
The core library is composed of couple classes -- mainly the \textit{Junction} and \textit{Layer} classes that define a basic magnetic component in the MTJ simulation and the \textit{Driver} class that contains definitions of various excitations that influence the magnetisation dynamics of the system. More detailed descriptions of the \textsc{cmtj} functionality, as well as implementation details, can be found in the library documentation. We describe the \textit{Layer}, \textit{Junction}, and \textit{Driver} interfaces in the following sections; details of additional functionality can be found in the documentation.
    
\subsection{Magnetic contributions}
The pivotal equation of the magnetic macrospin simulations is called Landau-Lifshitz-Gilbert-Slonczewski (LLGS) and takes the following form \cite{gilbert_classics_2004,ralph_spin_2008,slonczewski_current-driven_1996,nguyen_spinorbit_2021,ament_solving_2017}:
\begin{multline}
	  \frac{\textrm{d}\textbf{m}}{\textrm{dt}} = -\gamma_0 \textbf{m} \times \textbf{H}_{\mathrm{eff}} + \alpha_\textrm{G} \textbf{m}\times \frac{\textrm{d}\textbf{m}}{\textrm{dt}} \\
	 -\gamma_0 H_\textrm{FL}(\textbf{m} \times  \textbf{p}) -\gamma_0 H_\textrm{DL}(\textbf{m}\times\textbf{m}\times \textbf{p})
\label{eq:llg-sot}
\end{multline}    
where \(\mathbf{m} = \frac{\mathbf{M}}{M_s}\) is a normalised magnetisation vector, with \(M_\textrm{s}\) as the magnetisation saturation, $\alpha_\textrm{G}$ as the dimensionless Gilbert damping parameter, $\mathbf{p}$ is the polarisation vector, and \(\gamma_{0}\) is the gyromagnetic factor. The factors $H_\textrm{DL}$ and $H_\textrm{FL}$ are so-called damping- and field-like torque amplitudes, respectively, and generally take the form of STT or SOT, depending on the experiment. Each of those forms has a separate set of parameters and differs fundamentally in interpretation; for details, consult the documentation.

The effective field vector $\mathbf{H}_\mathrm{eff}$ is usually composed of various field contributions that, depending on the context of the simulation, may be added or disabled. The simulation package in question already provides a range of such contributions: 
\begin{equation}
    \mathbf{H}_\textrm{eff} = \mathbf{H}_\textrm{ext} + \mathbf{H}_\textrm{IEC} + 
    \mathbf{H}_\textrm{Oe} + 
    \mathbf{H}_\textrm{K} +
    \mathbf{H}_\textrm{demag} +
    \mathbf{H}_\textrm{dipole} + 
    \mathbf{H}_\textrm{th}^* +
    \mathbf{H}_\textrm{1/f}^*
\end{equation}
where each component corresponds, respectively, to the applied external field, interlayer exchange coupling (IEC), Oersted field, anisotropy field, demagnetising, dipole, thermal, and $1/f$ noise field \cite{perez_extending_nodate,voss_1f_1978}. Contributions marked with $^*$, if enabled in the simulation, require a stochastic solver, which is described in Sect.\ref{sec:solver}.
Each of the contributions that constitute $\mathbf{H}_\textrm{eff}$ may be varied over time using a driver system, as laid out in Sect.\ref{sec:drivers}.

\subsection{Simulation design}
The simulations in \textsc{cmtj} are conceptually divided into three levels of abstraction: single-layer simulations, multilayer simulations, and stacked device simulations. The first of those levels is designed to provide a basic interface of the ferromagnetic (FM) layer which, at low level, is described by a time-dependent magnetisation vector.

Multilayer simulation applications revolve most commonly around simple bilayer structures consisting of heavy metal (HM) and FM layers, or MTJs composed of FM layers, each separated by a thin tunnel barrier (TB). The non-ferromagnetic layers such as HM are naturally not simulated in the package, but their inclusion in the experiment has direct consequences in the simulation by its SOT property. For example, between two FM layers separated with TB there is IEC that varies with thickness \cite{katayama_interlayer_2006,koziol-rachwal_interlayer_2017}. Similarly, layers can be coupled in the multilayer system via a longer-range dipole interaction. HMs are typically simulated indirectly with the addition of STT or SOT terms \cite{liu_spin-torque_2012, miron_perpendicular_2011}. It is also possible to save computing time with the inclusion of pinned layers, experimentally obtained by the exchange-bias structure enforced by the presence of an antiferromagnet. In such a case, the solver is not run for a pinned/reference layer, but the effects of SOT or STT are still modelled via a constant reference vector.  
Finally, stack simulations, described in more detail in Sect.\ref{sec:stack}, refer to the composites of multilayer devices, which exhibit various electric coupling.

\subsection{Solver methods}
\label{sec:solver}
The core solver for most of the systems in \textsc{cmtj} is the Runge-Kutta 4 algorithm, as it balances decent convergence with speed. However, if the user specifies a temperature component for the system, \textsc{cmtj} switches to the stochastic solver, using the Euler-Heun method, and solves the Stratonovich form of the LLG equation \cite{bertotti_chapter_2009}:
\begin{equation}
    \mathrm{d}\mathbf{m}(t) = \mathbf{f}(\mathbf{m}(t), t)dt + \mathbf{g}(\mathbf{m}(t), t)\circ \mathrm{d}\mathbf{W_t}
    \label{eq:sde}
\end{equation}
where the non-stochastic part $\mathbf{f}(\mathbf{m}(t), t)$ is the LL form of the LLG equation, while the stochastic part, $\mathbf{g}(\mathbf{m}(t), t)$ contains thermal and other stochastic contributions:
\begin{equation}
    \mathbf{g}(\mathbf{m}_t, t)\circ\mathrm{d}\mathbf{W_t}  = 
    - \frac{\sigma\gamma}{1+\alpha^2}[\mathbf{m}\times\mathrm{d}\mathbf{W_t} + \alpha\mathbf{m}\times(\mathbf{m}\times\mathrm{d}\mathbf{W_t})]
    \label{eq:stochastic_part}
\end{equation}
The $\mathrm{d}\mathbf{W_t}$ is the random unit vector whose components are sampled from the normal distribution with zero mean, $\mathcal{N}(0, 1)$.

\subsection{Junction class system}
In \textsc{cmtj} each FM layer is by default independently solved, unless there is some coupling involved. To facilitate multilayer experiments, multiple FM layers may be combined into a \textit{Junction} with global control of certain magnetic and dynamic parameters. For instance, one can apply the same magnetic field drivers to all or selected layers of the multilayer system using the \textit{Junction} methods. This approach simplifies the code and provides a consistent API for both individual \textit{Layer} objects and whole \textit{Junction}.

\subsection{Driver class system}
\label{sec:drivers}
The dynamic pathway to excite any system in \textsc{cmtj} takes place through the driver system. The user can define any excitation in the form of a driver, an adequate \textit{Scalar} driver, or a vector (\textit{Axial}) driver. The latter are just compositions of the former along each dimension (the x, y, and z axes independently). The \textit{Scalar} driver is calculated at each time step, and influences the selected field contributions. For instance, one may define a sinusoidal driver and use it as a driving excitation of the anisotropy, leading to simulation of the VCMA effect. 
Likewise, the externally applied magnetic field can be added through the \textit{AxialDriver} class, by specifying field contributions along each axis independently.

\subsection{Stack class system}
\label{sec:stack}
As mentioned above, devices composed of layers can be further combined into \textit{Stack} objects. Each \textit{Stack} may be of type \textit{Parallel} or \textit{Series}, which determines the resistance calculation. The main consequence of composing a stack device type is the electrical coupling by a current passing through. The model for this kind of coupling follows the work of Taniguchi et al. \cite{taniguchi_mutual_2018}. For a parallel or series stack of two junctions, the current depends on the free magnetizations $\mathbf{m}$ and the pinning layers $\mathbf{p}$ of the junctions $i_{th}$ and $i+1_{th}$ as follows:
\begin{align}
    I(t) = I_0(t) + \chi I_0(t)(\mathbf{m_i} \cdot \mathbf{p}_i + \mathbf{m}_{i+1} \cdot \mathbf{p}_{i+1}) \quad \textrm{(series)}\\ 
    I(t) = I_0(t) + \chi I_0(t)(\mathbf{m_i} \cdot \mathbf{p}_i - \mathbf{m}_{i+1} \cdot \mathbf{p}_{i+1}) \quad \textrm{(parallel)}
\end{align}
where $I_0(t)$ represents the value of the uncoupled current. The coupling strength $\chi$ may be positive or negative, most of the time strictly much less than 1 in absolute value.

\section{Examples}
\label{sec:examples}
In this section, we focus on reproducing selected interesting experimental data and, where possible, comparing the simulation result with the experimental data. The examples are arranged by the level of complexity, starting from the simplest towards more advanced. 
We have chosen the following techniques to showcase in this section: standard M(H) and R(H) loops in Sect.\ref{sec:loops}, magnetoresistance (MR) based SD-FMR and PIMM in Sect.\ref{sec:GMR}, harmonic Hall voltage detection, angular variant in Sect.\ref{sec:harmonic-hall}, current-induced magnetisation switching (CIMS) in Sect.\ref{sec:switching}, electrical coupling of two MTJs in Sect.\ref{sec:electrical-coupling}.

All the examples presented in this section can be reproduced by running the \textsc{Jupyter} notebooks from the examples section on the \textsc{GitHub} repository of the \textsc{cmtj} package. 
\subsection{M(H) and R(H) loops}
\label{sec:loops}
\begin{figure}
    \centering
    \includegraphics[width=\linewidth]{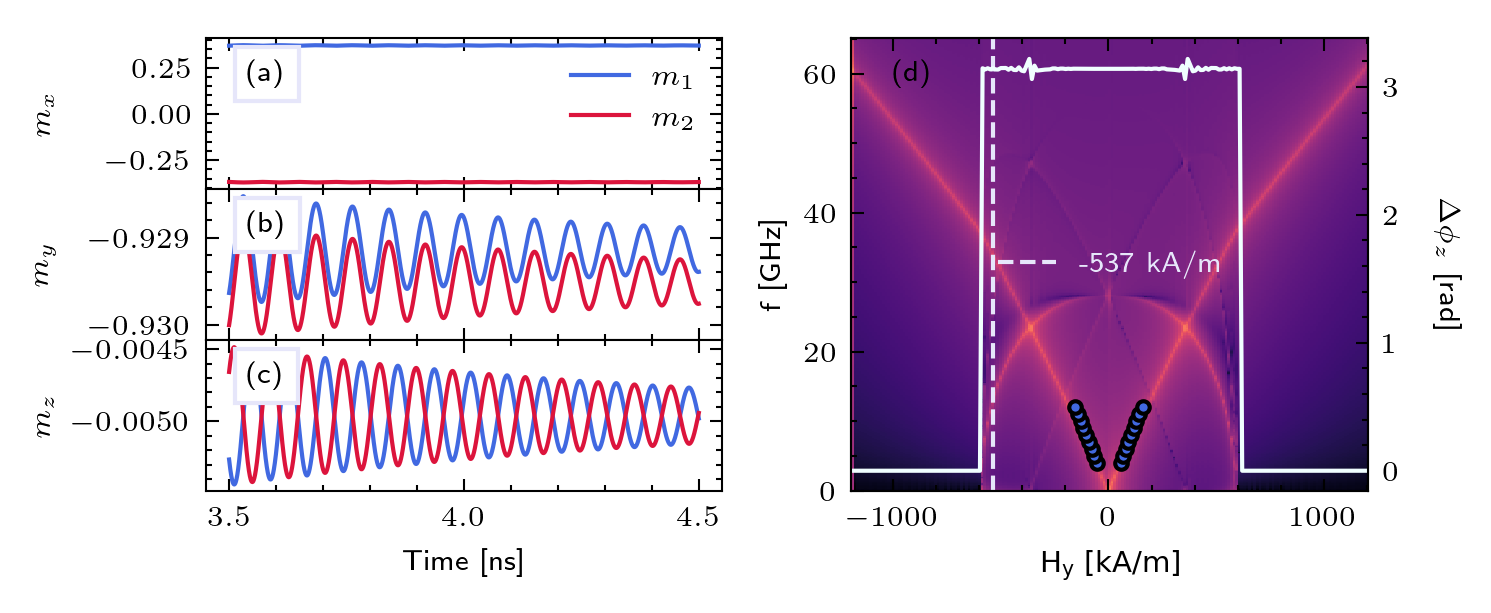}
    \caption{Selected magnetisation trajectory components (a-c) for a $H_\textrm{ext} = -537 \si{kA/m}$ for the Co(4nm)/Ru(0.53nm)/Co(4nm) system, marked with the dashed white line in (d). Colour denotes a given layer, blue is top and red is the bottom layer. (c) shows that the $m_z$ components of the two layers oscillate in antiphase. In Fig.\ref{fig:components}(d) we display Kittel dispersion relation obtained from SD-FMR measurements (described in Sect.\ref{sec:gmr-sd-fmr}), with experimental points given by blue dots, and phase difference $\Delta\phi_z$ between $m_z$ components of two layers (solid white line) across the external magnetic field range. The region between approximately -600 $\si{kA/m}$ and 600 $\si{kA/m}$, where the optical branch is visible, exhibits antiphase oscillation of $m_z$ components. Parameters taken from Tab.\ref{tab:parameters-pymag}.}
    \label{fig:components}
\end{figure}

The M(H) and R(H) are the basic magnetic characterisation methods that enable the determination of various material parameters of the multilayer system, such as magnetisation saturation, magnetic anisotropy, or magnetoresistance ratio. The M(H) loop simulates magnetometry measurements (such as a vibrating sample magnetometer (VSM) or magnetooptical Kerr effect (MOKE)), whereas the R(H) loops are indispensable in adjusting the magnetoresistance magnitude. Furthermore, by performing angular sweep simulations in different planes, one can also determine a type of magnetoresistance, because anisotropic MR (AMR), giant MR (GMR)/tunnelling MR (TMR) and spin-Hall MR (SMR) are characterised by different angular dependencies \cite{kim_spin_2016,cho_large_2015,rzeszut_biaxial_2018}. 
Other magnetoresistance flavours, such as those presented in, for example, the works by \cite{avci_unidirectional_2015,velez_hanle_2016}, may be easily added after the magnetisation dynamics has been computed by \textsc{cmtj}.
\begin{figure}
    \centering
    \includegraphics[width=\linewidth]{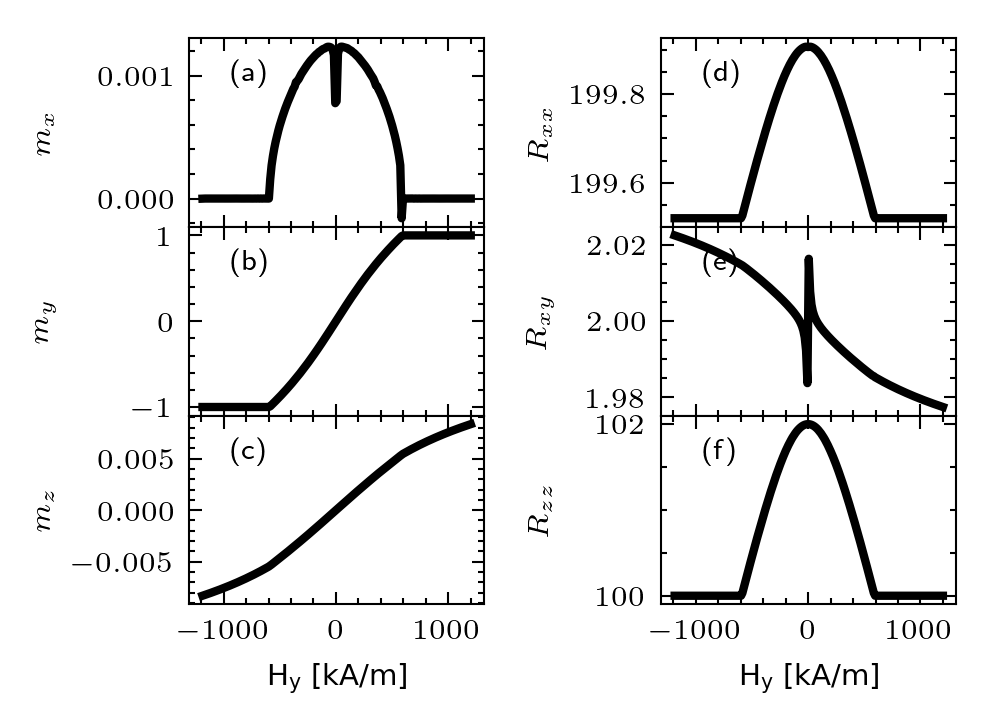}
    \caption{M(H) (a-c) and R(H) (d-f) loop of a Co/Ru/Co trilayer system described in the example. Parameters taken from Tab.\ref{tab:parameters-pymag}. The magnetic moment is normalised to unity, whereas the SMR, AMR and GMR magnitudes are set to -0.24 $\Omega$, -0.045 $\Omega$ and 2 $\Omega$, respectively (please note that the $m_y$ and $m_z$ components show very little variation with field H applied along y-axis).}
    \label{fig:loops}
\end{figure}
Magnetic field-dependent simulations of the Co(4 nm)/Ru(0.53 nm)/Co(4 nm) system are presented in Fig.\ref{fig:loops}(a-f). In this example, two FMs are strongly antiferromagnetically coupled through a Ru spacer \cite{mckinnon_spacer_2021}, resulting in a zero-net magnetic moment without an external magnetic field. Increasing H along the \textit{y} direction leads to a scissor-shaped magnetisation vector alignment, which is saturated at around 600 kA/m. This is made evident by the loop M(H) for the $m_y$ component illustrated in Fig.\ref{fig:loops}b), with the remaining, $m_x$ and $m_z$ having very small amplitudes. Furthermore, we show that, due to antiferromagnetic coupling (AFM), the two layers oscillate in antiphase, as shown in Fig.\ref{fig:components}.

\subsection{GMR system}
\label{sec:GMR}
A more complex example section begins with a study of a CoFe(2.1nm)/Cu(1.9–2.37nm)/CoFe(1nm)/NiFe(5nm) GMR sample with variable thickness of the Cu spacer layer. This system is characterised by oscillatory coupling with the spacer thickness well described in terms of the Rudermann-Kittel-Kasyua-Yosida (RKKY) interaction between the magnetisations of the reference and the free layers \cite{zietek_influence_2015}.
In the case of GMR/TMR, the magnetoresistance $R(\theta)$ is calculated according to the resistances in parallel ($R_P$) and antiparallel states ($R_{AP}$):
\begin{equation}
    R(\theta) = R_P + \frac{1}{2}(R_{AP} - R_{P})(1 - \mathbf{m}_{free}\cdot \mathbf{m}_{reference})
\end{equation}
where $\theta$ is the angle between the magnetisation vectors of the free and reference layer, $m_{free}$ is the magnetisation of the free layer and $m_{reference}$ is the magnetisation of the reference layer. Using the parameters for the system from the papers \cite{zietek_rectification_2015,zietek_influence_2015}, also summarised in Tab.\ref{tab:parameters-GMR}, we perform spin SD-FMR and PIMM simulations.

\subsubsection{SD-FMR}
\begin{figure}[h]
    \centering
    \includegraphics[width=\linewidth]{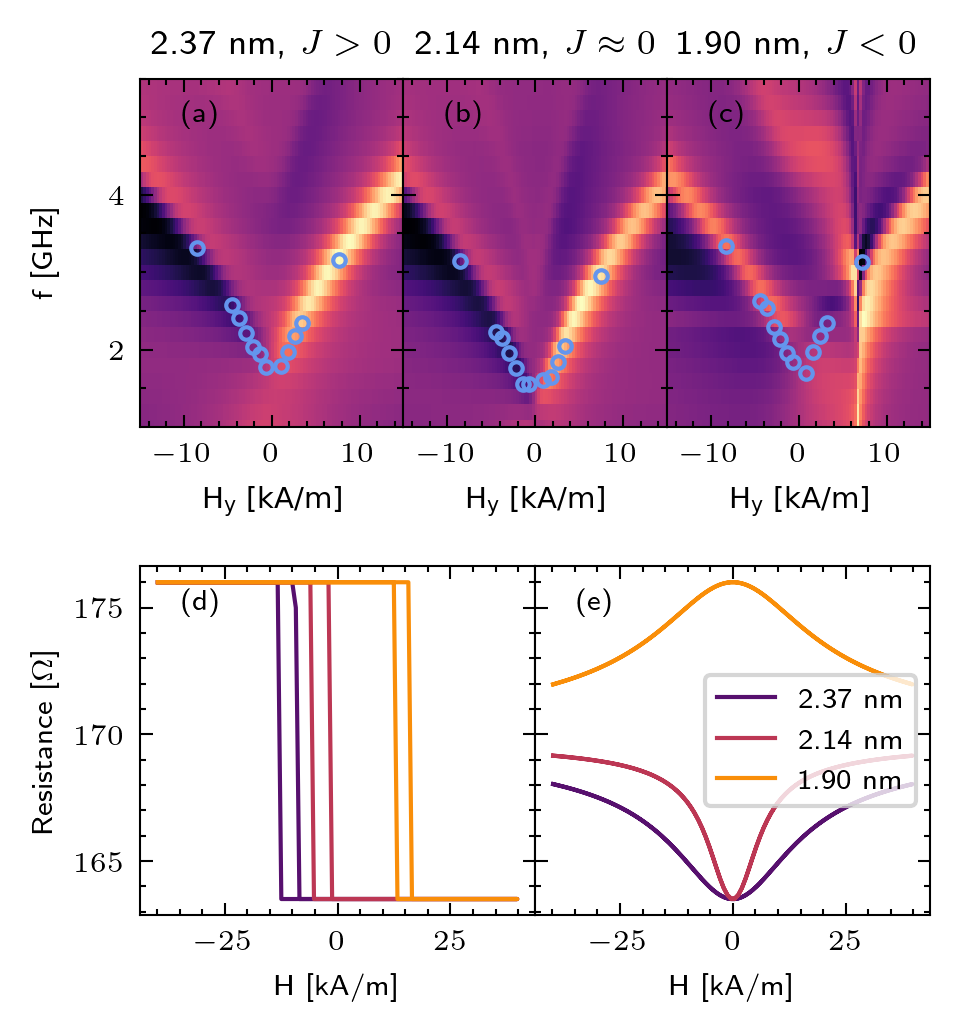}
    \caption{
    Kittel dispersion relations (a-c) for 3 different values of Cu thickness in CoFe(2.1nm)/Cu(1.9–2.37nm)/CoFe(1nm)/NiFe(5nm) sample, with the blue dots marking the experimental data. Resistance loops at  $\phi_H = 90^\circ$ (d) and at $\phi_H = 0^\circ$ (e) applied external magnetic field with $\phi$ denoting the polar angle. Parameters of the system may be found in Tab.\ref{tab:parameters-GMR}.}
    \label{fig:disp}
\end{figure}
In the simulation setup discussed, a representative of SD-FMR is modelled using the 2D map, $V_\textrm{mix}$ vs. $H$ and $f$, $V_{mix}(f, H)$, where $V_{mix}$ is the mixing voltage of $R(\theta)$ and $I_\textrm{AC}$ of a frequency $f$, under an applied external field $H$. The results are presented in Fig.\ref{fig:disp} with open blue circles as the experimental points for comparison.
\label{sec:gmr-sd-fmr}
Afterwards, we turn to the analysis of magnetisation dynamics in the GMR system by investigating its response using the SD-FMR method. In this experiment an AC current, $I_\textrm{AC}$, is supplied in a given frequency range, typically between 2 and 40 GHz, while performing a sweep of an external magnetic field.
As a result, the DC mixing voltage $V_\textrm{mix}$ is obtained as a function of both the frequency and the magnitude of the external magnetic field \cite{tulapurkar_spin-torque_2005}. This method allows for the calculation of the ferromagnetic resonance frequency for a given set of system parameters in the form of a Kittel dispersion relation. Furthermore, analysis of the shape of the simulated signal can be used to determine the spin torque values, in the current perpendicular to the plane STT \cite{sankey_measurement_2008} and current in the plane SOT geometries \cite{liu_spin-torque_2012}.
\begin{figure}[!h]
    \centering
    \includegraphics[width=\linewidth]{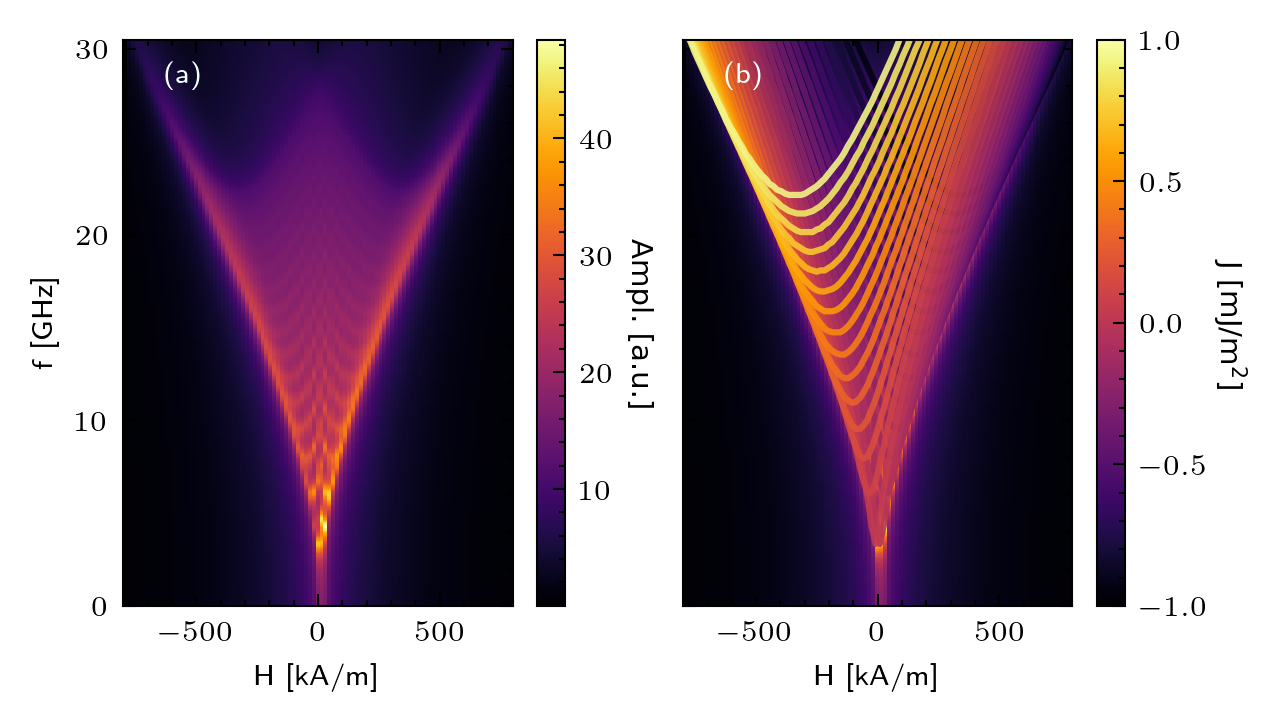}
    \caption{PIMM of the CoFe(2.1nm)/Cu(1.9–2.37nm)/CoFe(1nm)/NiFe(5nm) system described in Sec.\ref{sec:GMR}. (a) PIMM simulations for a range of IEC values in $(-1\, \si{mJ/m^2},1\, \si{mJ/m^2})$ combined into a single map. Here, the colour indicates the total frequency amplitude summed over all PIMM simulations. (b) illustrates the same PIMM, but with marked maximum amplitude resonance line for each $J$ value mapped accordingly to the colour scale. The colour scale attached represents different IEC values, from $-1\, \si{mJ/m^2} $ to $1\, \si{mJ/m^2} $. We observe the shift of the resonance curve towards the centre for smaller absolute IEC values.
    Note that at $H = 0$ the system still exhibits non zero oscillation.
    Parameters used for simulation can be found in Tab.
    \ref{tab:parameters-GMR}. Field applied along at $45^\circ$ angle between x and y axis.}
    \label{fig:pimm}
\end{figure}
\subsubsection{PIMM} 
Apart from the SD-FMR method described above, magnetisation dynamics can also be investigated by analysing the magnetisation response to picosecond magnetic field pulses. In the experiment, this time domain method is performed indirectly using PIMM \cite{silva_inductive_1999,serrano-guisan_inductive_2011,banasik_magnetic_2015}. The damped oscillatory response of the multilayer system to the short magnetic field pulse contains information on the characteristic resonance.
From the simulation perspective, PIMM is simulated as an analysis of the free oscillations contrary to the dynamics induced by the external alternating signal, as is the case for the FMR. If the FFT computation of the pulse response for each value of the applied magnetic field from the sweep is repeated, one can obtain a spectrogram representing the dispersion relation where each pixel represents the FFT magnitude for a given frequency and magnetic field.
For the GMR system, multiple PIMM simulations are conducted for a range of different IEC values, resulting in Fig.\ref{fig:pimm}(c). Clearly, as the magnitude of the IEC coupling increases, the resonance curves shift away from the zero field. This example shows the ease of predicting characteristic resonance frequencies in multilayer systems.

\subsection{Angular harmonic Hall voltage detection}
\label{sec:harmonic-hall}
\begin{figure}[!h]
    \centering
    \includegraphics[width=0.7\linewidth]{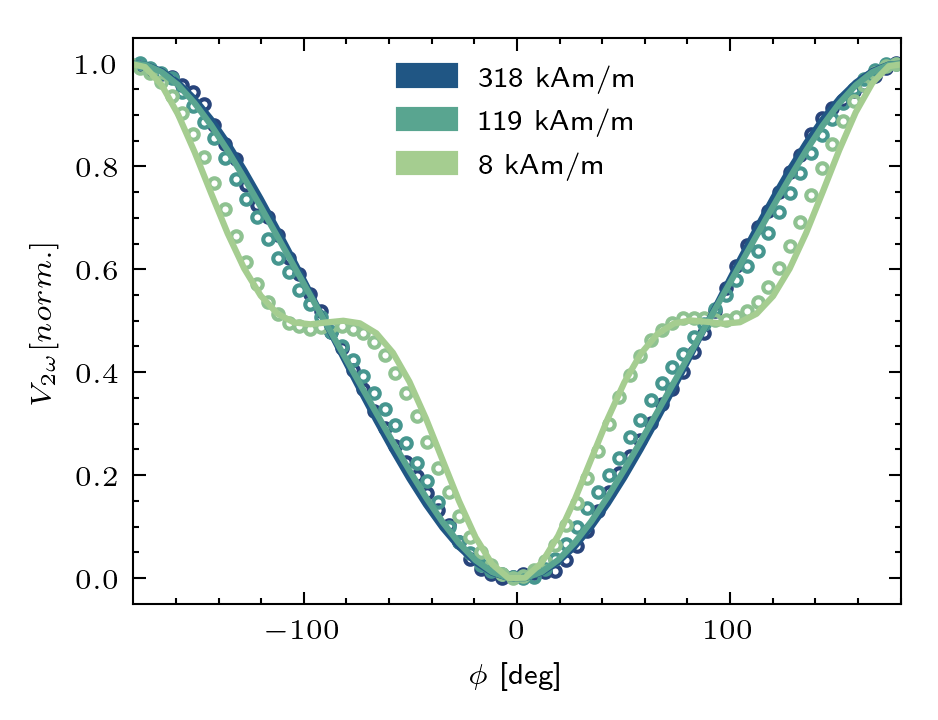}
    \caption{Second Harmonic Hall voltage detection in angular variation, dots represent experiment, lines demark the simulation data. The y-axis is normalised. Measurement data collected from Pt(4.3nm)/FeCoB(2nm) device.}
    \label{fig:harmonics}
\end{figure}
There are several experimental methods that allow the determination of SOT components, such as the SOT-FMR line shape \cite{liu_spin-torque_2011} and the line width analysis \cite{pai_spin_2012}, SOT switching current analysis \cite{hao_giant_2015}, or field-dependent harmonic Hall voltage techniques \cite{kim_layer_2013}. Another method, less susceptible to various artefacts, such as the anomalous Nernst effect, is called the angular-dependent harmonic Hall voltage \cite{nguyen_spinorbit_2021}. Our model has been thoroughly verified using the field-dependent method \cite{zietek_numerical_2022}, here we present a simplified angular variation of the harmonic Hall voltage detection. 
We show the simulation result, along with the experimental results for the corresponding Pt(4.3nm)/FeCoB(2nm) sample \cite{skowronski_angular_2021}, in Fig.\ref{fig:harmonics}. Each line is produced by sweeping with the azimuth angle $\phi \in [-180^\circ, 180^\circ]$ at the frequency excitation of the AC current $f$ and measuring the second harmonic output in the mixing voltage of the input current (for details, see \cite{skowronski_angular_2021}). Very good agreement is found between the experimental data and the simulation results.
The parameters for this system can be found in Tab.\ref{tab:parameters-harmonics}.

\subsection{CIMS}
\label{sec:switching}
\begin{figure}[h]
    \centering
    \includegraphics[width=0.55\linewidth]{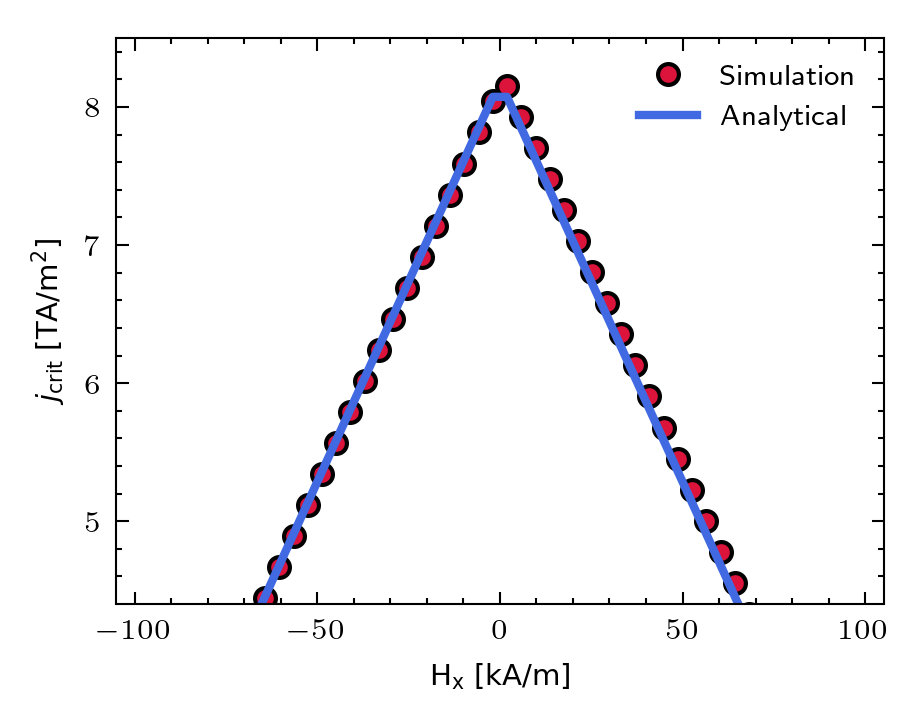}
    \caption{Analytical and simulated critical current density vs. external field $H_\textrm{x}$ for Pt(4nm)/Co(1nm)/MgO SOT device.}
    \label{fig:switching}
\end{figure}

Next, an example of the SOT-induced magnetisation switching in the HM/FM bilayer is discussed. The experimental data were obtained in the multilayer system: Pt(4nm)/Co(1nm)/MgO (A1) from the work by Grochot et al. \cite{grochot_current-induced_2021}. In this case, we reproduce the theoretical switching behaviour of the system using the field-like and damping-like SOT, and magnetic parameters obtained from the experiment. The result is shown in Fig.\ref{fig:switching}. We approximated the critical current analytically using a formula from Lee et al. \cite{lee_threshold_2013}:
\begin{equation}
    j_\textrm{sw}  \approx \frac{2 e \mu_0 M_\textrm{s}t_\textrm{FM}}{\hbar \theta_{SH}} (\frac{H_\textrm{K,eff}}{2} - \frac{H_x}{\sqrt{2}})
\end{equation}
where $e$ is the electron charge, $\mu_0$ is the magnetic permeability in vacuum, $M_\textrm{s}$ is the magnetisation saturation, $\hbar$ is the reduced Planck constant, $\theta_{SH}$ is the effective spin Hall angle, $H_\textrm{K,eff}$ is the effective perpendicular anisotropy value, and $H_x$ is the applied field along the $x$ direction.
For simulations, we used the trapezoidal impulse shape, with a rising and falling edge of 1 $\si{ns}$ and a flat edge of 3 $\si{ns}$. We also normalise the damping- and field-like torque magnitudes obtained from the experiment by the current density with which they have been measured. During the current density sweep, they will scale proportionally, giving the correct values of $H_\textrm{DL}$ and $H_\textrm{FL}$ at each step.

\subsection{Electrical coupling}
\label{sec:electrical-coupling}
\begin{figure}[!h]
    \centering
    \includegraphics[width=\linewidth]{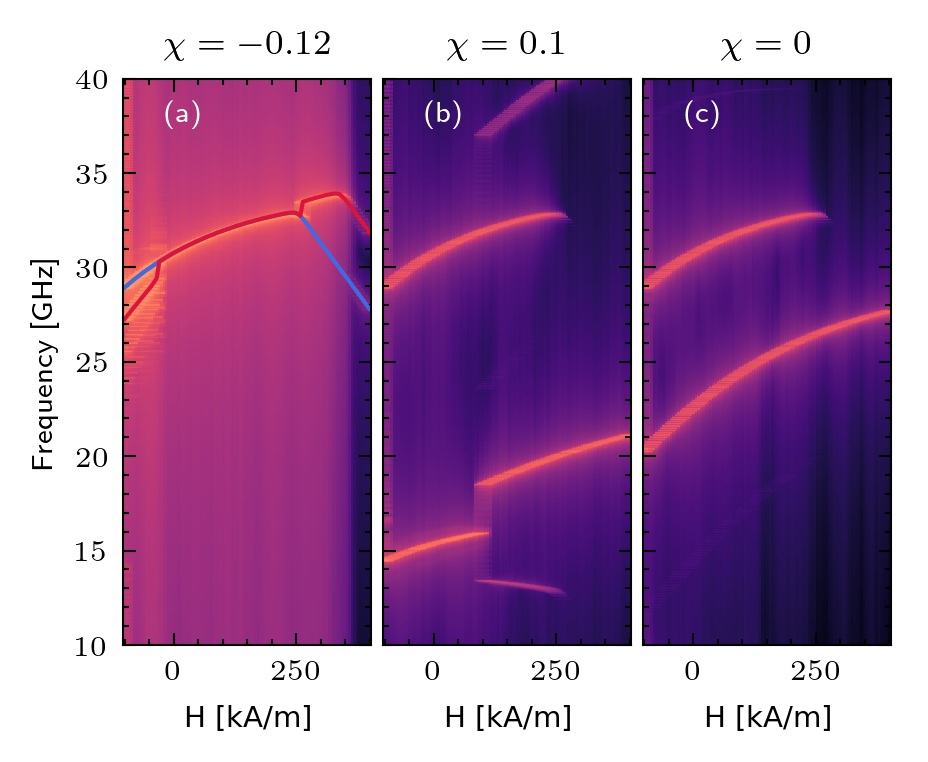}
    \caption{The electric synchronisation of two example MTJs, with electric coupling constant set to $\chi=-0.12$ (a) and $\chi=0.1$ (b). Solid blue and red lines in (a) indicate the primary resonance modes of individual MTJs from the stack. The Fig.\ref{fig:synch}(c) shows the same device with no coupling is present.}
    \label{fig:synch}
\end{figure}
Finally, this example illustrates how electric coupling can be simulated using \textsc{cmtj}. First, we created two MTJs with slightly different magnetic and resistance parameters, emulating experimental dispersion. Then, using the \textit{Stack} class interface (see Sect.\ref{sec:stack}), we couple them in a parallel connection, setting the coupling value $\chi$. Finally, we sweep the external magnetic field and measure the frequency response of the stack to constant current. The parameters of two systems have been collected in Tab.\ref{tab:parameters-electric-coupling}. In our case, we perform a sweep across a range of $H_\textrm{ext}$ values, applied at the azimuthal angle of $5^\circ$. Then, we measure the frequency response of the coupled system to the constant current density excitation. For larger values of the applied external magnetic field, and negative coupling constant $\chi$, we observe how two main resonance lines, each corresponding to a separate MTJ, converge towards a common resonance mode. Ultimately, around 250 $\si{kA/m}$, the two MTJs desynchronise and their main resonance lines separate again. The result of that electric synchronisation of two MTJs is depicted in Fig.\ref{fig:synch}(a), while Fig.\ref{fig:synch}(b-c) illustrate a situation with positive coupling coefficient and coupling disabled, respectively. This example shows that the software presented can also be used in more complex systems, for example, for the analysis of neural computing platforms \cite{romera_binding_2022}.

\subsection{Benchmarks}
The \textsc{cmtj} is meant to run also on personal computers, thus in Tab.\ref{tab:benchmark} we present example execution times that were measured on 2020 MacBook Pro with Apple Silicon M1 and 16 GB RAM. The \textsc{cmtj} library includes additional helper functions that allow for easy parallelism if the simulation setup allows it. Running certain simulations in parallel without considering initial position effects may produce drastically different results. For instance, while sweeping with an external magnetic field, we can treat each next step as a slight perturbation relative to the previous one, and thus we can use the final step magnetisation from $H_i$ as the starting position for $H_{i+1}$ simulation. To compensate for this, for example in the parallel setup, one can extend the simulation with an initial relaxation period to stabilise the magnetisation first.

\begin{table*}[!ht]
\centering
\begin{tabular}{r|c|c|c|c}
Benchmark             & Steps {[}$10^6${]} & Distributed & Time {[}s{]} & Steps/s {[}$10^3$/s{]} \\ \hline
M(H), R(H), PIMM      & 50                    & NO          & 42                     & 1190                       \\
CIMS                  & 187.5                 & YES         & 355                    & 528                        \\
GMR VSD               & 1872                  & YES         & 367                    & 5101                       \\
GMR PIMM              & 360                   & YES         & 315                    & 1143                       \\
Stack synchronisation & 120                   & NO          & 62                     & 1935                      
\end{tabular}
\caption{The execution times and the number of steps per experiment are given in the examples section (Sect.\ref{sec:examples}). The numbers summarise all sweeps over parameters, fields, and frequencies, and thus represent the total measured time required to reproduce the experiment. The timing was done on a MacBook Pro 2020 with an Apple M1 processor and 16 GB of RAM. Selected examples benefitted from parallel computation to show the ease of use of the included package parallelism method based on the \textsc{Python} multiprocessing library. Some experiments, such as, for instance, CIMS (15 ns), involve more parameter scanning than individual simulation steps, so they are less efficient than simulations with longer in-simulation time, such as GMR VSD (60 ns).}
\label{tab:benchmark}
\end{table*}

\section{Conclusions}
In this article, we presented \textsc{cmtj}, fast modelling software for macrospin simulation of magnetic heterostructures. Its core is grounded in the LLGS equation, and the package is capable of calculating both the static and dynamic characteristics of spintronic devices. In the spirit of the modern modular development approach, suggested, for example, in \cite{locatelli_spin-torque_2014, camsari_modular_2015}, \textsc{cmtj} expands its usability to connect multiple spintronic structures using the \textit{Junction} or \textit{Stack} system. Furthermore, we show that the results obtained from \textsc{cmtj} agree well with the experiment and that the calculated simulation derivatives are shown to be consistent in both the static and dynamic regimes. 

\section{Acknowledgements}
We thank P. Ogrodnik for a fruitful and insightful discussion. The research project is partly supported by the \textit{Excellence initiative – research university} programme for the AGH University of Science and Technology.
T.S. and K.G. were supported by the National Science Centre, Poland, Grant No. UMO-2016/23/B/ST3/01430 (SPINORBITRONICS).

\clearpage
\newpage
\begin{appendices}
\section{\textsc{PyMag} example}
\begin{figure*}[!tbp]
    \centering
    \includegraphics[width=\textwidth]{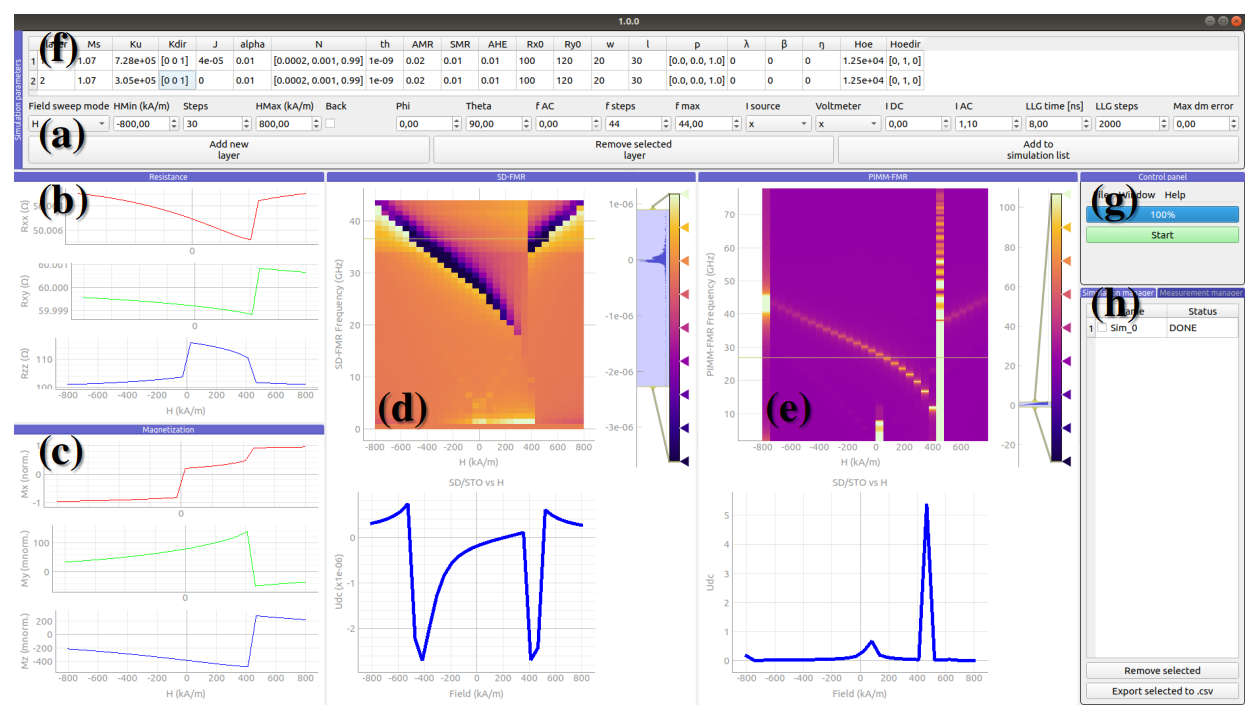}
    \caption{View of \textsc{PyMag} UI (a) Control panel, (b) sample layer and stimulus settings panel, (c) R-H plots, (d) M-H plots, (e) map of PIMM results, (f) map of SD-FMR results, (g) simulations management panel, (h) experimental or external simulation management panel.}
    \label{fig:PyMagView}
\end{figure*}
\textsc{PyMag} is a supplementary GUI programme that incorporates certain experimental procedures using the \textsc{cmtj} interface, such as M(H), R(H), PIMM or SD-FMR in a user-friendly interface. 
In this section, we discuss step by step how to reproduce a simple simulation with \textsc{PyMag} software. \textsc{PyMag} allows for the definition of the material parameters and stimuli of the spintronics device, the simulation process, the management of the simulation list, previewing and analysis of the simulation results, the management of loaded experimental data, the export of selected results, and the saving and loading of the entire simulation state. \textsc{PyMag} interface was based on open source libraries: \textsc{PyQt5} and \textsc{pyqtgraph} \textsc{Python}. In addition, \textsc{PyMag} allows the user to compare their experimental data with the simulated results using the \textit{Experiment management} tab.

The structure parameters of each layer of the device must be defined in the table; see Fig.\ref{fig:PyMagView}(f). Each row corresponds to a single ferromagnetic layer and contains a variety of parameters. For convenience we put them here, placing the units expected by \textsc{PyMag} in brackets: magnetisation saturation $\mu_0 M_\mathrm{s} (\si{T})$, uniaxial anisotropy $K_\mathrm{u} (\si{J/{m^3}}$), interlayer exchange coupling, first order $J (\si{J/{m^2}})$ and second order $J_2 (\si{J/{m^2}})$, geometrical properties (thickness, width , length  $d (\si{m})$, $w (\si{\mu m})$, $l (\si{\mu m})$, respectively), anisotropic magnetoresistance ($\mathrm{AMR}$), spin Hall magnetoresistance ($\mathrm{SMR}$), anomalous Hall effect ($\mathrm{AHE}$) (all in $\Omega$), and the following dynamic parameters: damping factor $\alpha$, field-like torque $H_\mathrm{FL}$ ($\si{A/m}$), damping-like torque $H_\mathrm{DL}$ ($\si{A/m}$). The Oersted field excitation $H_\textrm{Oe}$ is given in the table in $\si{1/m}$, as it is multiplied by the current value entered in the stimulus section. The following parameters are entered as vector quantities (each being a vector of dimension 3): $\mathbf{K}_\mathrm{u dir}$, the direction of uniaxial anisotropy, $\mathbf{N}_\textrm{demag}$, the diagonal of the demagnetisation tensor, $\mathbf{p}$, the polarisation vector in the SOT torque, and $\mathbf{H}_\textrm{Oedir}$, the direction of the Oersted field.

For a basic, demonstrative example we have chosen a trilayer structure, with two ferromagnetic layers separated with a spacer layer, Co/Ru/Co, which exhibits a quadratic interlayer exchange coupling, producing optical and acoustic lines in the FMR spectrum. 
The multilayer parameters used in the simulations are grouped in Tab.\ref{tab:parameters-pymag}.

Then we move on to the definition of a stimulus, which can be defined in the \textit{Stimulus} tab.
In \textsc{PyMag} we follow a physical spherical coordinate system that designates $\theta$ as the polar angle (measured from the positive z-axis), and $\phi$ as the azimuth angle. In our software, the external field excitation may be swept either by magnitude or by any of those two angles. Furthermore, there is a DC ($I_\mathrm{DC}$) and AC ($I_\mathrm{AC}$) current that can be supplied in a selected frequency range described by three parameters: $f_\mathrm{min}$, $f_\mathrm{max}$, $f_\mathrm{steps}$). 
For the current example, we keep $I_\textrm{DC}$ fixed and sweep the magnetic field along the y-axis between -600 and 600 kA/m.

Finally, the precision of the simulation is governed by the total simulation time in nanoseconds: $\textrm{LLG}_\mathrm{time}$ and the number of integration steps ($\textrm{LLG}_\mathrm{steps}$) in the RK4 method. We find that, for most cases, the integration step should not be greater than $10^{-12}$ to ensure the correctness and proper convergence of the simulation. However, this depends on multiple factors. For instance, large IEC coupling, as in the case of the example, may lead to very complex trajectories that require much shorter integration times to converge properly. We propose setting 500000 steps per $5\si{ns}$, which corresponds to the integration step of $10^{-14}\si{s}$. When inspecting the \textit{Convergence} tab, the user can conveniently observe the regions of good and poor convergence, defined as a running difference in magnitude of $\frac{d\mathbf{m}}{dt}$.

The control panel (Fig.\ref{fig:PyMagView}(g-h)) is used to monitor the simulation status. The user can load or save selected simulations, displaying the experimental data directly on the simulated runs. The progress bar informs the user about the stage of the run, and the stop and start buttons control the simulation flow. To run a simulation, we click on \textit{Run simulation button} in the control panel. The graphs in the visualisation tabs are populated online as the simulation progresses. 
In \textsc{PyMag}, the results of this aggregation can be found in the \textit{PIMM} tab (Fig.\ref{fig:PyMagView}(e))

\clearpage
\section{Layer parameters}
\subsection*{Parameters explained}
\label{sec:params}
\begin{table}[h!]
\centering
\begin{tabular}{r|l}
Parameter                   & Meaning                              \\ \hline
$\mu_0 M_\textrm{s}$        & Magnetisation saturation             \\ \hline
$J$                         & IEC value                            \\ \hline
$J_\textrm{quad}$           & IEC value (quadratic term)           \\ \hline
$K_{u}$                     & Anisotropy value                     \\ \hline
$\mathbf{K}_{dir}$          & Anisotropy axis                      \\ \hline
$\alpha_\textrm{G}$         & Gilbert damping parameter            \\ \hline
$\mathbf{N}_\textrm{demag}$ & diagonal of demagnetisation tensor   \\ \hline
$t_\textrm{FM}$             & thickness of a FM layer              \\ \hline
$w$                         & width of a sample                    \\ \hline
$l$                         & length of the sample                 \\ \hline
$H_\textrm{DL}$             & magnitude of the damping-like torque \\ \hline
$H_\textrm{FL}$             & magnitude of the field-like torque   \\ \hline
$\mathbf{p}$                & polarisation vector                  \\ \hline
$\chi$                      & electric coupling constant           \\ \hline
$\lambda$                   & Slonczewski spacer layer parameter   \\ \hline
$\eta$                      & spin polarisation efficiency        
\end{tabular}
\caption{Explanation of the key parameters used in the simulations.}
\label{tab:parameter-expl}
\end{table}

\clearpage
\subsection*{Example 1 -- Co(4nm)/Ru(0.53nm)/Co(4nm)}
\begin{table}[h!]
\centering
\begin{tabular}{rrl}
\multicolumn{1}{r|}{Parameter}                      & \multicolumn{1}{r|}{Value}                     & Unit         \\ \hline
\multicolumn{3}{c}{Magnetic parameters}                                                                             \\ \hline
\multicolumn{1}{r|}{$\mu_0 M_\textrm{s}^{1,2}$}           & \multicolumn{1}{r|}{1.65}                      & $\si{T}$     \\
\multicolumn{1}{r|}{$K_{u}^{1,2}$}                  & \multicolumn{1}{r|}{1050}                      & $\si{J/m^3}$ \\
\multicolumn{1}{r|}{$\mathbf{K}_{dir}^{1,2}$}          & \multicolumn{1}{r|}{{[}1, 0, 0{]}}             & -            \\
\multicolumn{1}{r|}{$J^1$}                          & \multicolumn{1}{r|}{-1.78}                  & $\si{mJ/m^2}$ \\
\multicolumn{1}{r|}{$J_{quad}^1$}                   & \multicolumn{1}{r|}{-0.169}                 & $\si{mJ/m^2}$ \\
\multicolumn{1}{r|}{$\alpha_\textrm{G}^{1,2}$}      & \multicolumn{1}{r|}{0.005}                     & -            \\
\multicolumn{1}{r|}{$\mathbf{N}_\textrm{demag}^{1,2}$} & \multicolumn{1}{r|}{{[}0.001, 0.000, 0.998{]}} & -            \\ \hline
\multicolumn{3}{c}{Dimensions}                                                                                      \\ \hline
\multicolumn{1}{r|}{$t_\textrm{FM}^1$}              & \multicolumn{1}{r|}{4}                         & $\si{nm}$    \\
\multicolumn{1}{r|}{$t_\textrm{FM}^{2(*)}$}              & \multicolumn{1}{r|}{3.99}                      & $\si{nm}$    \\
\multicolumn{1}{r|}{$w^{1,2}$}                      & \multicolumn{1}{r|}{20}                        & $\si{um}$    \\
\multicolumn{1}{r|}{$l^{1,2}$}                      & \multicolumn{1}{r|}{30}                        & $\si{um}$    \\ \hline
\multicolumn{3}{c}{Resistance parameters}                                                                           \\ \hline
\multicolumn{1}{r|}{$\textrm{AMR}^{1,2}$}           & \multicolumn{1}{r|}{-0.045}                    & $\si{Ohm}$   \\
\multicolumn{1}{r|}{$\textrm{SMR}^{1,2}$}           & \multicolumn{1}{r|}{-0.24}                     & $\si{Ohm}$   \\
\multicolumn{1}{r|}{$\textrm{AHE}^{1,2}$}           & \multicolumn{1}{r|}{-2.7}                      & $\si{Ohm}$   \\
\multicolumn{1}{r|}{$\textrm{GMR}^{1,2}$}           & \multicolumn{1}{r|}{2}                      & $\si{Ohm}$   \\
\multicolumn{1}{r|}{$\textrm{R}_\textrm{xx0}^{1,2}$}       & \multicolumn{1}{r|}{100}                       & $\si{Ohm}$   \\
\multicolumn{1}{r|}{$\textrm{R}_\textrm{xy0}^{1,2}$}       & \multicolumn{1}{r|}{1}                         & $\si{Ohm}$  
\end{tabular}
\caption{Parameter table for the example trilayer Co(4nm)/Ru(0.53nm)/Co(4nm) system. The superscript denotes the layer id. If there is no parameter for the corresponding layer id, the value is 0 by default. $(^*)$ We distinguish between the thicknesses of the upper and lower Co layers due to the effect of the interfacial magnetic dead layer. The parameter explanation may be found in the previous section, Tab.\ref{tab:parameter-expl}}
\label{tab:parameters-pymag}
\end{table}

\clearpage
\subsection*{Example 2 -- CoFe(2.1nm)/Cu(1.9–2.37nm)/CoFe(1nm)/NiFe(5nm)}
\begin{table}[h!]
\centering
\begin{tabular}{rrl}
\multicolumn{1}{r|}{Parameter}                      & \multicolumn{1}{r|}{Value}         & Unit          \\ \hline
\multicolumn{3}{c}{Magnetic parameters}                                                                  \\ \hline
\multicolumn{1}{r|}{$\mu_0 M_\textrm{s}^{1}$}             & \multicolumn{1}{r|}{1.03}          & $\si{T}$      \\
\multicolumn{1}{r|}{$\mu_0 M_\textrm{s}^{2}$}             & \multicolumn{1}{r|}{1.65}          & $\si{T}$        \\
\multicolumn{1}{r|}{$K_{u}^{1}$}                    & \multicolumn{1}{r|}{0.8}           & $\si{kJ/m^3}$ \\
\multicolumn{1}{r|}{$K_{u}^{2}$}                    & \multicolumn{1}{r|}{$1$}           & $\si{TJ/m^3}$ \\
\multicolumn{1}{r|}{$\mathbf{K}_{dir}^{1,2}$}          & \multicolumn{1}{r|}{{[}1, 0, 0{]}} & -             \\
\multicolumn{1}{r|}{$\alpha_\textrm{G}^{1,2}$}      & \multicolumn{1}{r|}{0.024}         & -             \\
\multicolumn{1}{r|}{$\mathbf{N}_\textrm{demag}^{1,2}$} & \multicolumn{1}{r|}{{[}0, 0, 1{]}} & -             \\ \hline
\multicolumn{3}{c}{Dimensions}                                                                           \\ \hline
\multicolumn{1}{r|}{$t_\textrm{FM}^1$}              & \multicolumn{1}{r|}{2.1}      & $\si{nm}$     \\
\multicolumn{1}{r|}{$t_\textrm{FM}^2$}              & \multicolumn{1}{r|}{6}             & $\si{nm}$     \\ \hline
\multicolumn{3}{c}{Resistance parameters}                                                                \\ \hline
\multicolumn{1}{r|}{$\textrm{R}_\textrm{P}$}               & \multicolumn{1}{r|}{163.5}           & $\si{Ohm}$    \\
\multicolumn{1}{r|}{$\textrm{R}_\textrm{AP}$}              & \multicolumn{1}{r|}{176}           & $\si{Ohm}$   
\end{tabular}
\caption{Parameter table for the example GMR system. The superscript denotes the layer id.}
\label{tab:parameters-GMR}
\end{table}

\clearpage
\subsection*{Example 3 -- Angular harmonics for Pt(4.3nm)/FeCoB(2nm)}
\begin{table}[h!]
\centering
\begin{tabular}{rrl}
\multicolumn{1}{r|}{Parameter}                      & \multicolumn{1}{r|}{Value}         & Unit          \\ \hline
\multicolumn{3}{c}{Magnetic parameters}                                                                  \\ \hline
\multicolumn{1}{r|}{$\mu_0 M_\textrm{s}$}             & \multicolumn{1}{r|}{1.2}           & $\si{T}$      \\
\multicolumn{1}{r|}{$K_{u}$}                    & \multicolumn{1}{r|}{1.49}          & $\si{kJ/m^3}$ \\
\multicolumn{1}{r|}{$\mathbf{K}_{dir}$}          & \multicolumn{1}{r|}{{[}0, 0, 1{]}} & -             \\
\multicolumn{1}{r|}{$\alpha_\textrm{G}$}      & \multicolumn{1}{r|}{0.003}         & -             \\
\multicolumn{1}{r|}{$\mathbf{N}_\textrm{demag}$} & \multicolumn{1}{r|}{{[}0, 0, 1{]}} & -             \\ \hline
\multicolumn{3}{c}{Dimensions}                                                                           \\ \hline
\multicolumn{1}{r|}{$t_\textrm{FM}$}              & \multicolumn{1}{r|}{2}      & $\si{nm}$     \\
\multicolumn{1}{r|}{$w$}                              & \multicolumn{1}{r|}{10}            & $\si{\mu m}$  \\
\multicolumn{1}{r|}{$l$}                              & \multicolumn{1}{r|}{30}            & $\si{\mu m}$  \\ \hline
\multicolumn{3}{c}{Resistance parameters}                                                                \\ \hline
\multicolumn{1}{r|}{$\textrm{AHE}$}                 & \multicolumn{1}{r|}{1.15}          & $\si{Ohm}$    \\
\multicolumn{1}{r|}{$\textrm{SMR}$}                 & \multicolumn{1}{r|}{-0.125}        & $\si{Ohm}$    \\
\multicolumn{1}{r|}{$\textrm{AMR}$}                 & \multicolumn{1}{r|}{$-10^{-4}$}    & $\si{Ohm}$    \\ \hline
\multicolumn{3}{c}{Torque parameters}                                                                    \\ \hline
\multicolumn{1}{r|}{$H_\textrm{DL}$}                & \multicolumn{1}{r|}{1054}          & $\si{A/m}$    \\
\multicolumn{1}{r|}{$H_\textrm{FL}$}                & \multicolumn{1}{r|}{-164}          & $\si{A/m}$    \\
\multicolumn{1}{r|}{$\mathbf{p}$}                   & \multicolumn{1}{r|}{{[}0, 1, 0{]}} & -            
\end{tabular}
\caption{Parameter table for the angular harmonics example, Pt/CoFeB sample.}
\label{tab:parameters-harmonics}
\end{table}

\clearpage
\subsection*{Example 4 -- Electric coupling}
\begin{table}[h!]
\centering
\begin{tabular}{rrl}
\multicolumn{1}{r|}{Parameter}                    & \multicolumn{1}{r|}{Value}         & Unit          \\ \hline
\multicolumn{3}{c}{Magnetic parameters}                                                                \\ \hline
\multicolumn{1}{r|}{$\mu_0 M_\textrm{s}^1$}             & \multicolumn{1}{r|}{1.6}           & $\si{T}$      \\
\multicolumn{1}{r|}{$\mu_0 M_\textrm{s}^2$}             & \multicolumn{1}{r|}{1.76}          & $\si{T}$      \\
\multicolumn{1}{r|}{$K_{u}^1$}                    & \multicolumn{1}{r|}{70}            & $\si{kJ/m^3}$ \\
\multicolumn{1}{r|}{$K_{u}^2$}                    & \multicolumn{1}{r|}{100}           & $\si{kJ/m^3}$ \\
\multicolumn{1}{r|}{$\mathbf{K}_{dir}^{1,2}$}     & \multicolumn{1}{r|}{{[}0, 0, 1{]}} & -             \\
\multicolumn{1}{r|}{$\alpha_\textrm{G}$}          & \multicolumn{1}{r|}{0.005}         & -             \\
\multicolumn{1}{r|}{$\mathbf{N}_\textrm{demag}$}  & \multicolumn{1}{r|}{{[}0, 0, 1{]}} & -             \\ \hline
\multicolumn{3}{c}{Resistance parameters}                                                              \\ \hline
\multicolumn{1}{r|}{$\textrm{R}_\textrm{P}^{1}$}  & \multicolumn{1}{r|}{100}           & $\si{Ohm}$    \\
\multicolumn{1}{r|}{$\textrm{R}_\textrm{P}^{2}$}  & \multicolumn{1}{r|}{110}           & $\si{Ohm}$    \\
\multicolumn{1}{r|}{$\textrm{R}_\textrm{AP}^{1}$} & \multicolumn{1}{r|}{200}           & $\si{Ohm}$    \\
\multicolumn{1}{r|}{$\textrm{R}_\textrm{AP}^{2}$} & \multicolumn{1}{r|}{220}           & $\si{Ohm}$    \\ \hline
\multicolumn{3}{c}{Torque parameters}                                                                  \\ \hline
\multicolumn{1}{r|}{$\lambda$}                    & \multicolumn{1}{r|}{0.69}          & -             \\
\multicolumn{1}{r|}{$\eta^{1,2}$}                 & \multicolumn{1}{r|}{1}             & -             \\
\multicolumn{1}{r|}{$\mathbf{p}^{1,2}$}           & \multicolumn{1}{r|}{{[}1, 0, 0{]}} & -             \\ \hline
\multicolumn{3}{c}{Stack parameters}                                                                   \\ \hline
\multicolumn{1}{r|}{$\chi$}                       & \multicolumn{1}{r|}{-0.12}          & -             \\
\multicolumn{1}{r|}{$j$}                          & \multicolumn{1}{r|}{60}            & $\si{MA/m^2}$
\end{tabular}
\caption{Parameter table for the electric coupling system of two MTJs connected in parallel. The current density was set to 60 \si{MA/m^2}. The parameters partially taken from \cite{skowronski_microwave_2019}, with the MTJ composition of CoFeB(2.5nm)/MgO (1nm)/CoFeB(1.8nm).}
\label{tab:parameters-electric-coupling}
\end{table}
\end{appendices}

\newpage
\bibliography{references.bib}

\end{document}